# RFID-based Solutions for Smarter Healthcare

**Cristina Turcu\*. Cornel Turcu.\*\***

*\* Stefan cel Mare University of Suceava, Suceava, Romania
(e-mail: \*cristina.turcu@usv.ro, \*\*cornel.turcu@usv.ro)*

**Abstract:** This paper proposes the application of RFID technology in healthcare industry based on its increased functionality, high reliability, easy-to-use capabilities and low cost. After a brief presentation of RFID technologies and their applications, the paper describes an RFID-based system that can provide efficient facilities to allow essential information management for emergency care across hospital boundaries. This system performs RFID-based identification of the patients, querying and retrieving medical data from various existing healthcare information systems, as well as storing and giving the most clinically significant information to the clinicians. Also, the system allows identifying and tracking RFID- tagged objects in order to provide new quality services for the mobility of objects.

## 1. INTRODUCTION

In 2000, the United States (U.S.) National Institute of Medicine issued the report entitled: "To Err Is Human, Building a Safer Health System", which drew attention on the spreading problem of often preventable medical errors throughout the U.S. hospitals (Kohn, Corrigan & Donaldson, 2000). The report emphasizes that each year more than 98,000 deaths and 770,000 adverse drug events in the U.S. are caused by preventable medical errors. These are significant numbers considering the fact that the U.S. has the highest healthcare expenditure in the world.

This paper aims to show how Radio Frequency IDentification (RFID) technologies can be used for building smarter healthcare by reducing errors, improving patient safety, and optimizing business processes.

In the second section, we briefly introduce RFID technology and some of its applications in healthcare. The third section presents an RFID-based system that could be used in smarter healthcare initiative. The conclusion summarizes the main achievements of this paper.

## 2. RFID TECHNOLOGIES IN HEALTHCARE

### 2.1 RFID Background

Radio Frequency IDentification (RFID) technologies are wireless Automatic Identification and Data Capture (AIDC) technologies that allow the automatic identification of living or non-living entities, collecting data about these entities and storing this data on computer systems. RFID technology is similar to a well-known and widely used AIDC technology which is barcode technology. Although barcodes offer some advantages over RFID, (especially their low cost), there are a number of characteristics particular to RFID which makes this technology superior to barcodes in terms of (1) non optical proximity communication, (2) information density, (3) two- way communication ability and (4) multiple simultaneous reading (the reading of more than one item at once) (Roberts, 2006).

The basic RFID system architecture has two components: contactless electronic tags and an RFID reader. The RFID tag is used to store unique identification data and other specific information whereas the RFID reader allows reading and writing these tags. An RFID tag is attached to or embedded in the individual that is to be identified. Tags fall into three categories: active (battery-powered), passive (the reader signal is used for activation) or semi-passive (battery-assisted, activated by a signal from the reader).

RFID systems require software, network and database components that should enable information flow from tags to the organisation's information infrastructure, where the information is processed and stored. Systems are application-specific (OECD, 2008).

On a worldwide level, various RFID applications are employed across a wide and rapidly expanding range of industries. Thus, RFID applications offer solutions for: 1) logistical tracking and tracing, 2) production, monitoring and maintenance, 3) product safety, quality and information, 4) access control and tracking and tracing of individuals, 5) loyalty, membership and payment, 6) healthcare, 7) sport, leisure and household, 8) public services.

The next section briefly presents some of the real-world applications of RFID technologies in healthcare.

### 2.2 RFID Applications in Healthcare

According to various studies and reports, RFID technologies provide numerous solutions for the main areas of healthcare industry. Thus, e.g., the Kalorama Information report, *The Global Market for RFID in Healthcare*, considers the market for RFID opportunities in the healthcare industry, focusing on five market segments: 1) pharmaceutical/blood product distribution and tracking, 2) patient/medical staff identification and tracking, 3) medical asset tracking and locating, 4) implantable device RFID use, 5) other areas (including medical documents and patient records) (Kalorama, 2010).

Below are some examples of RFID applications in healthcare:

*Supply chain applications*. RFID solutions optimise supply chain management (SCM) in the medical/ pharmaceutical industry by ensuring improved visibility throughout the supply chain.

*Patient care applications*. This may include improved positive patient identification through the use of RFID tags to reduce the number of incidents harmful to patients (e.g., wrong drug or blood type), and the number of cases of mismatching between the baby and the mother. Also, real-time tracking of a patient's location in the hospital results in higher safety conditions and improved bed placement.

*Inventory management applications*. Not requiring line-of-sight scanning, RFID technology improves inventory management, allowing for a reduction of the time necessary to perform an accurate inventory and a decrease in overall inventory costs, etc.

*Asset management*. RFID solutions improve traceability of assets, allowing localisation of a mobile asset at all times, which reduces leases and the amount of time medical staff spend searching for equipment.

*Home-based healthcare*. Application domains include various telemedicine solutions that consider RFID applied for in-patient or out-patient.

*RFID sensor healthcare applications*. This may include wireless sensors to monitor patient temperature, Parkinson's disease, post-surgery awakening, etc.

These applications require tagging of assets (medical equipment) and/or actors (doctors, nurses, patients, etc.) etc. Furthermore, RFID readers are placed in important locations within the hospital, healthcare provider institution, etc.

Various studies (e.g., BRIDGE project (BRIDGE, 2007)) estimate a significant increase of RFID use in healthcare and pharmaceutical industry in the coming years (Table 1).

**Table 1. Forecast for RFID use in healthcare and pharmaceutical industry**

|   | 2007 | 2012 | 2017 | 2022 |
|---|---|---|---|---|
| Total RFID tags (in Millions) | 8 | 352 | 1720 | 6740 |
| - on hospital assets | 2 | 98 | 190 | 320 |
| - on laboratory samples | 1 | 8 | 30 | 40 |
| - on drugs | 5 | 246 | 1500 | 6380 |
| Locations with RFID readers | 110 | 2.770 | 11.900 | 40.600 |
| Total number of RFID readers | 180 | 12.600 | 70.200 | 208.000 |

Once the cost of RFID has fallen dramatically within the past few years and continues to decrease, researchers estimate that healthcare industry will benefit from RFID in a number of other applications. Next an RFID-based case study for healthcare is presented.

## 3. CASE STUDY

### 3.1 General Presentation

Today's Romanian medical sector has not taken full advantage from all the achievements of information systems. Patient-related information is scattered among various medical units, the patients' charts have no standardized form or content and are seldom complete or up-to-date; moreover, if need be, they cannot be accessed online by the medical staff. Another issue of healthcare system is related to supply chain and inventory management, such as, theft, counterfeiting, etc.

Considering these major inconveniencies, an RFID-based system for the distributed medical field could prove to be a viable solution. Thus, we propose a system, here named vITALIS (Internet of Things based Health Information System), that allows the management of the information related to any object or person, unique identifiable: patient, medical staff, medical equipment, etc.

This system enables real-time patient identification and monitoring, by employing the latest Radio Frequency Identification and multi-agent technologies. It ensures collaborative problem solving in distributed environment and provides communication infrastructure with multi-point connections to the medical information within the system. Thus, key medical information about the patient, such as drug allergies, blood type, etc., is stored on the individual's medical card (which is, in fact, a passive RFID tag). This card provides quick access to the information regarding the actual health state of a patient and helps the medical staff take the best decisions, especially in case of emergency. Thus, by using this system, even when the patient is unconscious, incoherent or unable to speak, emergency medical staff can read patient's card to quickly check blood type, current treatments, medical and allergy history.

VITALIS proposes a dual solution, a referential and a non-referential approach. Thus, the considered card may also store some data that provide secure access to the patient's electronic health care records kept in an electronic health record (EHR) system, located elsewhere. Hence, authorized staff can get additional health information about the patient, like past medical history, etc. This RFID-based system could be employed to ensure positive patient identification (PPI) within a hospital. Furthermore, it extends patient identification across hospital boundaries, through the use of a specialised agent that implements specific information sharing protocol. Hence, the system can integrate with existing medical information systems. Moreover, the application running on RFID-enabled mobile devices gives healthcare providers the information and capabilities they need wherever and whenever they need them. Thus, this RFID-based software system is an open-loop RFID application that could function across hospital boundaries.

Furthermore, the proposed system involves attaching RFID tags to objects such as, mobile medical assets, medical laboratory items, sensors, etc. The RFID tag may store some key information about the tagged-object and some data that allows accessing detailed information about the object from the specified location. Also, by using RFID-sensor tags for temperature monitoring, the system allows easily measuring and monitoring the temperature of tagged people or objects and saving the related data to corresponding entity records.

*3.2 System Actors*

Health information is currently stored in many different locations, such as with caregiver or other provider, physicians' offices, hospitals and laboratories databases, organizational and governmental Web sites, as well as public health databases. The considered system proposes a decentralized approach that should leave medical information in the database of the institutions that have a direct relationship with the patient, rather than moving or replicating it to a giant central server. But the system facilitates health information exchange in order to give clinicians full medical information about a patient. Interoperability with other predefined components in the healthcare sector is a facility provided by this system. As Figure 1 illustrates, there are considered independent components, integrated in the overall IT management of a hospital (e.g. Hospital Information System-HIS, Medical Information System-MIS) or the medical devices operated by the emergency medical team. The capability of reading out data from patient's card represents an important privacy issue and, therefore, extensible security protocols are required.

*3.3 System Benefits*

The presented system proposes an RFID-based approach for patient identification and ensures the gathering of medical information stored in other various medical information systems. Therefore, it can easily relate to any EHR system, already installed in today's medical establishments. Thus, it eliminates the import of all patients' electronic health records into other EHR systems. Furthermore, no new training is required for members of the medical staff on using a new information system in order to manage their patients' medical records. One of the flaws of present healthcare industry is rendered by the existence of duplicate patient medical tests, e.g. tests that have already been done in the previous one or two years. The presented system rectifies this flaw as is allows authorized medical staff to retrieve patient health information, such as medical test results, stored in other caregiver databases, in other health information systems, located elsewhere. This system ensures significant cost savings and patient care quality improvement.

Here are some of the major benefits offered by the presented system:

- clinical benefits - helping to improve patient safety by reducing medical errors; this can be done by: a) providing the medical staff a quick access to key medical data stored on patients' cards; b) allowing the secure distribution upon request of a whole range of patient-related information (such as patient medical histories).
- administrative benefits: a) considerably reducing the consumption of paper used for keeping hardcopy documents; b) efficiently managing medical information (such as, the results of medical investigations) by connecting health care providers; c) real-time identifying and tracking tagged-objects; d) enhancing inventory management; e) helping improving stock control.

Furthermore, this system could be easily extended by adding supplementary hardware or software modules. Also, the system allows the implementation of new features in order to integrate the system in the new vision of the so-called Internet of Things (IoT). Simply said, IoT can be considered as a shift in paradigm: "from anytime, anyplace connectivity for anyone, we will now have connectivity for anything" (ITU, 2005). According to (Botterman, 2009), IoT integrates "things having identities and virtual personalities operating in smart spaces using intelligent interfaces to connect and communicate within social, environmental, and user contexts". Thus, the Internet of Things infrastructure allows connections between different entities, such as, human beings (patients, medical staff, etc.), medical devices, wireless sensors, mobile robots, etc. The presented system proposes a solution for extending IoT in the healthcare domain through the use of RFID technology.

## 4. CONCLUSIONS

Many of the errors occurring in healthcare are related to the lack of availability of important medical information about the patient. According to researchers, the use of information technology (IT) and electronic medical records (EMR) holds promise for improving the quality of information transfer and is essential to patient safety (Bates & Gawande, 2003). Furthermore, new advances in information and computing technologies will lead to similarly dramatic changes in the healthcare environment. A vast and multilayered infrastructure of ubiquitous computing technologies and applications is emerging. The current widespread deployment of cell phones, laptops, Wi-Fi, Bluetooth, personal digital assistants (PDAs), and various forms of sensing devices based on digital and radio frequency identification (RFID) technologies penetrate the healthcare environment.

This paper considers some real-world applications of RFID technologies in healthcare area and briefly describes the benefits of RFID technology for healthcare applications. Also, a solution based on RFID, multi-agent and ontology technologies is considered. Thus, we propose tagging people and objects with RFID tags. Hence, an RFID-based card storing key medical information about a patient is introduced as a bearer electronic document. The presented system provides instant access to emergency medical information about the patient and offers secure access to electronic health records, even under the conditions of a patient's mobility. Hence, this system enables the holder of the card to receive high-quality medical aid in emergencies, but is not limited to these cases. The system allows identifying and tracking tagged objects in order to provide new quality services for the mobility of objects.

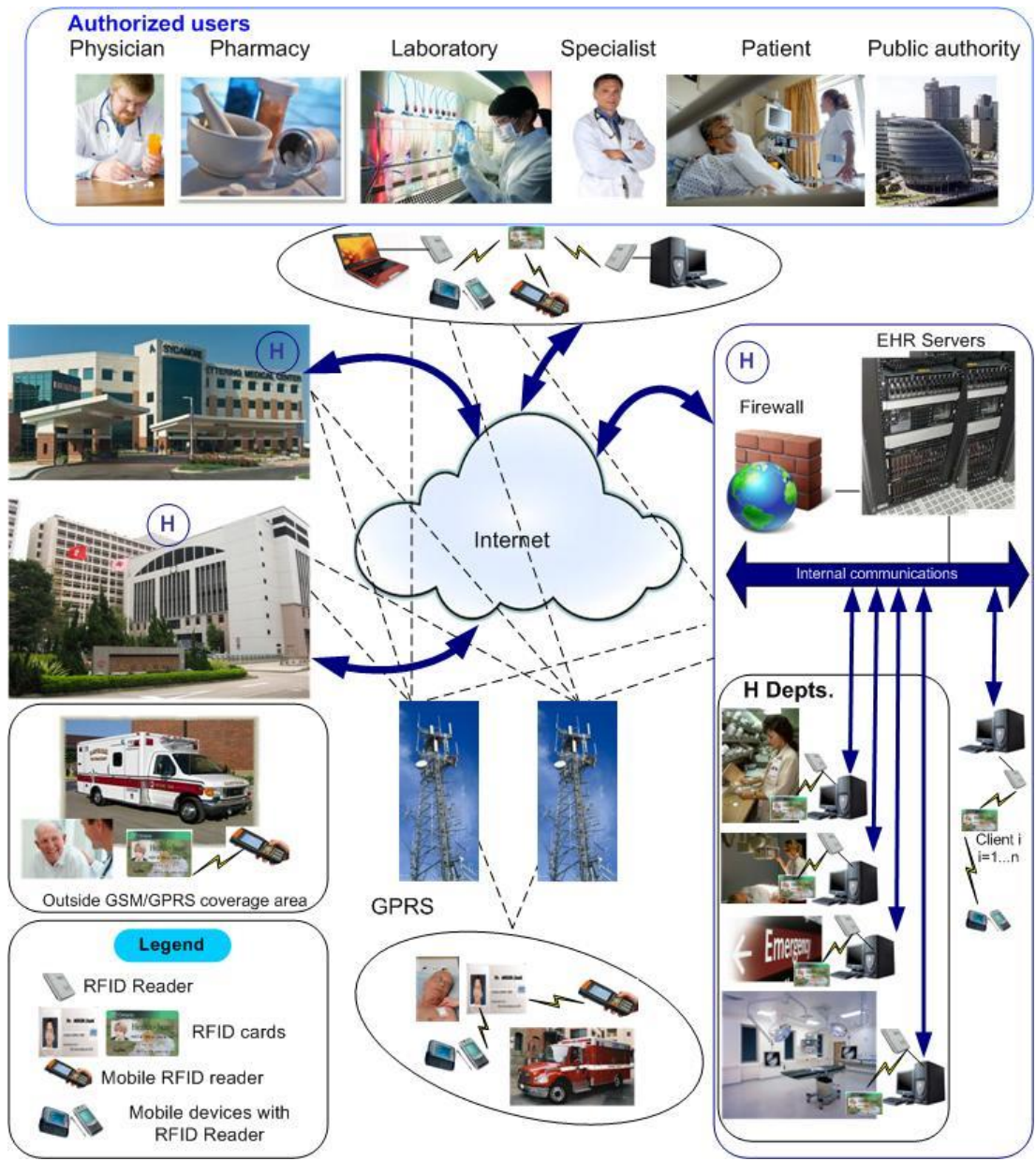

Fig 1. VITALIS Actors

ACKNOWLEDGMENTS

Research for this paper was supported by projects "PRiDE - Progress and development through post-doctoral research and innovation in engineering and applied sciences (contract no. POSDRU/89/1.5/S/57083", project co-funded from European Social Fund through Sectorial Operational Program Human Resources 2007-2013).